\documentstyle[prb,aps]{revtex}
\begin{document}

\input{epsf}
\draft

\twocolumn[\hsize\textwidth\columnwidth\hsize\csname @twocolumnfalse\endcsname
  
\title{Blume-Emery-Griffiths Model in a Random Crystal Field\cite{cnpq}}

\author{ \sc { N.\ S.\ Branco }} 
\address{Departamento de F\'{\i}sica - Universidade Federal de Santa
Catarina\\ 
88040-900, Florian\'opolis, SC  - Brazil; e-mail:
nsbranco@fisica.ufsc.br \\}

\date{\today}

\maketitle

\begin{abstract}

       We study the Blume-Emery-Griffiths 
model in a random crystal field in two and three dimensions, through a 
real-space renormalization-group approach and a mean-field approximation,
respectively. According to the two-dimensional renormalization-group
calculation, non-symmetry-breaking first-order phase 
transitions are eliminated and symmetry-breaking discontinuous transitions are
replaced by continuous ones, when disorder is introduced.
On the other hand, the mean-field calculation predicts that first-order 
transitions
are not eliminated by disorder, although some changes are introduced in the
phase diagrams. We make some comments on the consequences of a degeneracy
parameter, which may be relevant in martensitic transitions.
 	
\end{abstract}

 
\vskip2pc]

\section{Introduction}
                  
     The Blume-Emery-Griffiths (BEG) model is a spin-1 Ising model, originally 
proposed to study $^3$He-$^4$He mixtures\cite{BEG}. Later, it was
used in the description of a variety of different physical phenomena: 
multicomponent fluids \cite{exp1}, microemulsions 
\cite{exp2}, semiconductor alloys \cite{exp3}, electronic 
conduction models \cite{exp4}, etc. Its hamiltonian reads:
\begin{equation}
   {\cal  H} = - J \sum_{<i,j>} S_i S_j - K \sum_{<i,j>} S_i^2 S_j^2
   + \sum_i \Delta_i S_i^2, \;\;  
   \label{hamil} 
\end{equation}
where the first two sums are over all nearest-neighbor 
pairs on a lattice, the last one is over all sites 
and $S_i = \pm 1,0$. $J$ is the exchange parameter, $K$ is the
biquadratic interaction and $\Delta_i$ is a site-dependent crystal field
($\Delta_i=\Delta$ for the pure model).
The phase diagram of the model presents first-order
and continuous phase transitions and, for $K<0$, a rich variety of 
multicritical points \cite{ber1,ber2}.

	Nevertheless, some systems were found to be better described
by a disordered BEG model, as, for instance, conventional shape memory alloys 
\cite{barce} and fluid mixtures on disorder materials (like aerogel, for
example) \cite{mar1,mar2}.
From the theoretical point of view, on the other hand, it has been argued that 
randomness may have drastic consequences on multicritical behavior 
\cite{Imry,berker}. In
two dimensions, for instance, any infinitesimal amount of disorder
supresses non-symmetry-breaking first-order phase transitions and
replaces symmetry-breaking first-order phase transitions by continuous ones.
The effect of disorder on three-dimensional systems is different:
first-order phase transitions only disappear at a finite amount of
randomness \cite{berker}. This behavior has been observed in some
models \cite{berker,branco,cardy,falicov}. 

	In order to study the effects of disorder on its phase-diagram, 
we study the BEG  model in a random crystal field (henceforth called RBEG 
model) given by the probability distribution:
\begin{equation}
{\cal P}(\Delta_i) = r \; \delta(\Delta_i+\Delta) +
(1-r) \; \delta(\Delta_i-\Delta) \label{distri}
\end{equation}
It is worthy stressing that the exact
form of the disorder is not relevant to the overall {\it qualitative}
consequences on the phase diagram. If randomness is chosen to be
in the interactions $J$ or $K$, the qualitative effects will be the same
(in what concerns first-order phase transitions).
This is due to the fact that, even if the initial disorder is on the bonds 
(interactions
$J$ or $K$), a scale transformation will propagate this disorder to the 
crystal field
term, which will act just like field randomness on the coexistence boundary.
Moreover, the exact form of the probability distribution is not relevant, 
either; we have performed calculations with other distributions and they lead
to the same qualitative picture as the one found in this work.

	Finally, we would like to mention that, to the best of our knowledge,
the BEG model in a random crystal field has not been studied so far. Previous
studies concentrated on the random Blume-Capel model
\cite{mar1,mar2,branco,mf1,mf2}, which has a simpler phase-diagram than the
BEG model's.

	The remainder of this paper is organized as follows. In section
II we outline the mean-field approximation we use and discuss the results, 
in section III we present the real-space renormalization-group (RSRG) 
calculation (expected to hold for two-dimensional systems), 
and in the last section we summarize our main conclusions and comment on the
influence of a degeneracy parameter $p$ on the critical behavior.  

\section{Mean-field Calculation} 
	
	We chose an ordinary mean-field approximation to study the
three-dimensional system. The procedure is rather usual and we refer the
reader to Ref. \onlinecite{bachmann}, where a detailed discussion of the
method is done.

	However, we would like to stress that the mean-field approximation
we use is equivalent
to a model where the interaction is of infinite-range, i.e., each
spin interacts with every other spin in the system. This will have
explicit consequences on the phase diagram and we will return to this
point later.

	Most of the information about the phase diagram is numerically 
calculated
but some analytical results can be obtained. Among them, we can find the
ground state for any values of $J$, $K$, $\Delta$ and $r$.

	It is possible to show that the ground state magnetization, $m_0$, for 
$d\equiv \Delta/zJ > 0$, where $z$ is the coordination 
number of the lattice, is given by [results for $\Delta<0$ can be inferred 
from the mapping $(r,\Delta) \leftrightarrow (1-r,-\Delta)$]:
\begin{equation}
   m_0 = 1 - (1-r) \;  \theta\left[ d - \left( k+1 \right) \left( \frac{1+r}{2}
   \right)
   \right] ,
\end{equation}
where $k \equiv K/J$ and $\theta[x]$ is the step function, such that
$\theta[x]=0$ or $1$ for $x<0$ or $x>0$ respectively. The ferromagnetic
phase $O_1$ (see figures in this subsection), with $m_0=1$, is stable for 
$d \leq d_c = \left( k+1 \right) \left( \frac{1+r}{2} \right)$, while
for $d \geq d_c$ the ground state is such that $m_0=r$ (denoted $O_2$ in 
our figures). Note that, except for $r=0$, the ground state is always
ordered; this is a consequence of the simple mean-field approximation
we used (we will return to this point below).

	One can obtain the continuous transition line exactly,
by expanding $\Phi_{min}$ in powers of the magnetization
$m$ and taking the coefficient of $m^2$ equal to zero:
\begin{equation}
	t_c = 2 \left( \frac{1-r}{2 + e^{-k} e^{d/t_c}} +
	\frac{r}{2 + e^{-k} e^{-d/t_c}} \right), 
\end{equation}
where $t_c \equiv k_B T_c / z J$.
More specifically, note that, for $d \gg 1$, the value of the critical
temperature is $t_c = r$. So, for any value
of $r \neq 0$, the critical line between the paramagnet and the $O_2$ phases
extends to $d = \infty$ (see figures in this subsection). This is not
the expected behavior for a cubic lattice, for the following reason. The RBEG 
model for $d = \infty$ is equivalent to the site-diluted spin-1/2 Ising
model, since for $\Delta=\infty$, a $+\Delta$ crystal
field acting on a given site forces that site to be in the $S=0$ state 
(absent), while a $-\Delta$ field forces the site to be either in the state 
$S=1$ or in the state $S=-1$ (both represent a present site). Thus, only for
high enough $r$ an infinite cluster of $S=\pm 1$ states will form and will be 
able to sustain order. 
Exactly at $r=r_c$, there is such an infinite
cluster but its critical temperature is zero. Therefore, the critical 
parameter $d_c$ should only reach infinity for $r \geq r_c$. However, the 
simple mean-field analysis we made leads to $r_c=0$, since
it is equivalent to a model with infinite-range interactions.
In some cases \cite{mar1,mf2}, more elaborated mean-field-like
procedures were applied to the Blume-Capel model in a random crystal field.
Briefly, the consequence of these approaches is that the transition line
between $O_2$ and $D$ phases does not extend to $d = \infty$
for all values of $r$. All other results are similar to the ones obtained
with our simple mean-field approximation.
We note in advance that the approach we used for the two-dimenisonal model
leads to a finite value of $r_c$, as expected on physical grounds.

	We have already pointed out that $t_c(\Delta=\infty)$ does not 
depend on $K$; this comes
from the mapping between the RDBEG model and the site-diluted spin-1/2 Ising
model. The $S=0$ states (absent sites) play no role in the dynamics of 
the model and the present sites can only be in the states $S=1$ or $S=-1$;
thus, the biquadractic interaction, $K$, is irrelevant in this limit. If, 
for instance, the probability
distribution ${\cal P}(\Delta_i) = r \; \delta(\Delta_i) +
(1-r) \; \delta(\Delta_i-\Delta) $ is used, the $\Delta = \infty$ limit
will be equivalent to the site-diluted {\it spin-1} Ising model; then,
$t_c(\Delta=\infty)$ will depend on $K$. Note that the
discussion in this paragraph applies to the two-dimensional case as well.

\begin{figure}
\epsfxsize=6.5cm
\begin{center}
\leavevmode
\epsffile{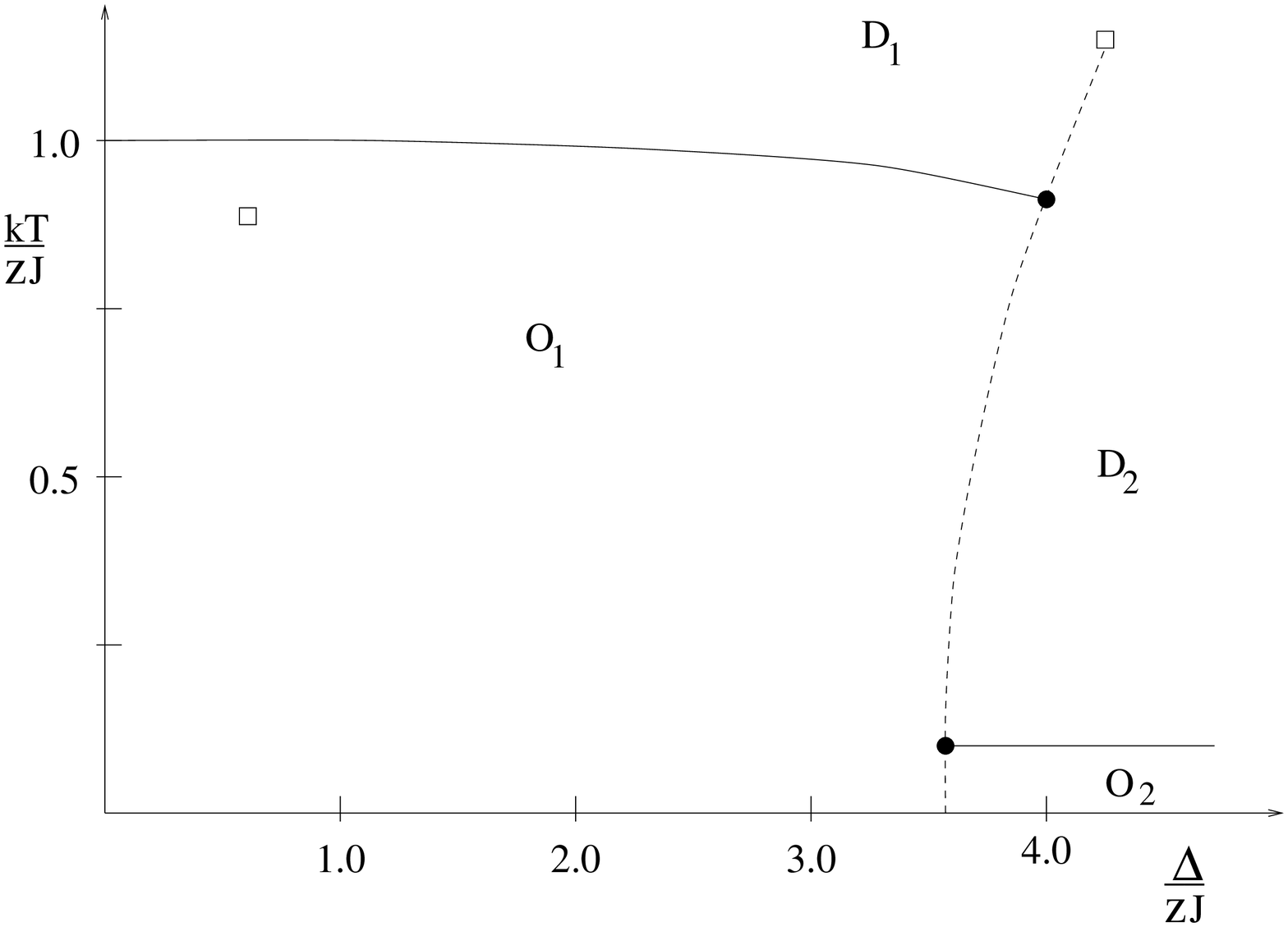}
\caption{Mean-field phase diagram for $K/J=5$ and $r=0.1$.
Filled circles and the open square stand for critical end points 
and a critical point,
respectively. Continuous (dashed) lines represent continuous (first-order)
transitions. The phases are: ordered with $m=1$ $(O_1)$, ordered with
$m=r$ $(O_2)$, disordered with $q>1/2$ $(D_1)$, and disordered with $q<1/2$ 
$(D_2)$.}
\label{k5r01} 
\end{center}
\end{figure}

\begin{figure}
\epsfxsize=6.5cm
\begin{center}
\leavevmode
\epsffile{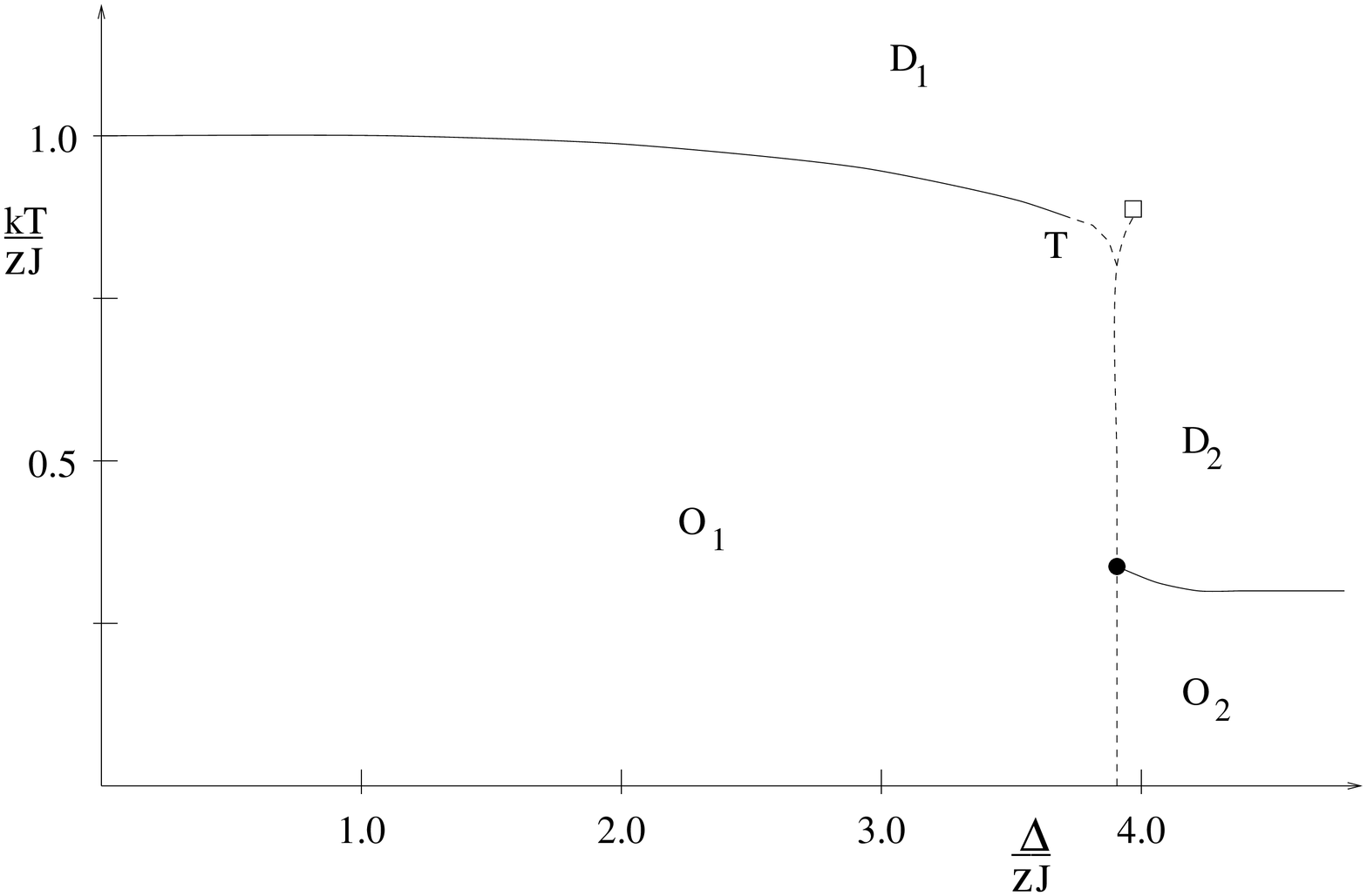}
\caption{Mean-field phase diagram for $K/J=5$ and $r=0.3$.
Same conventions as in Fig.~\ref{k5r01}; $T$ stands for 
tricritical points.}
\label{k5r03} 
\end{center}
\end{figure}

\begin{figure}
\epsfxsize=6.5cm
\begin{center}
\leavevmode
\epsffile{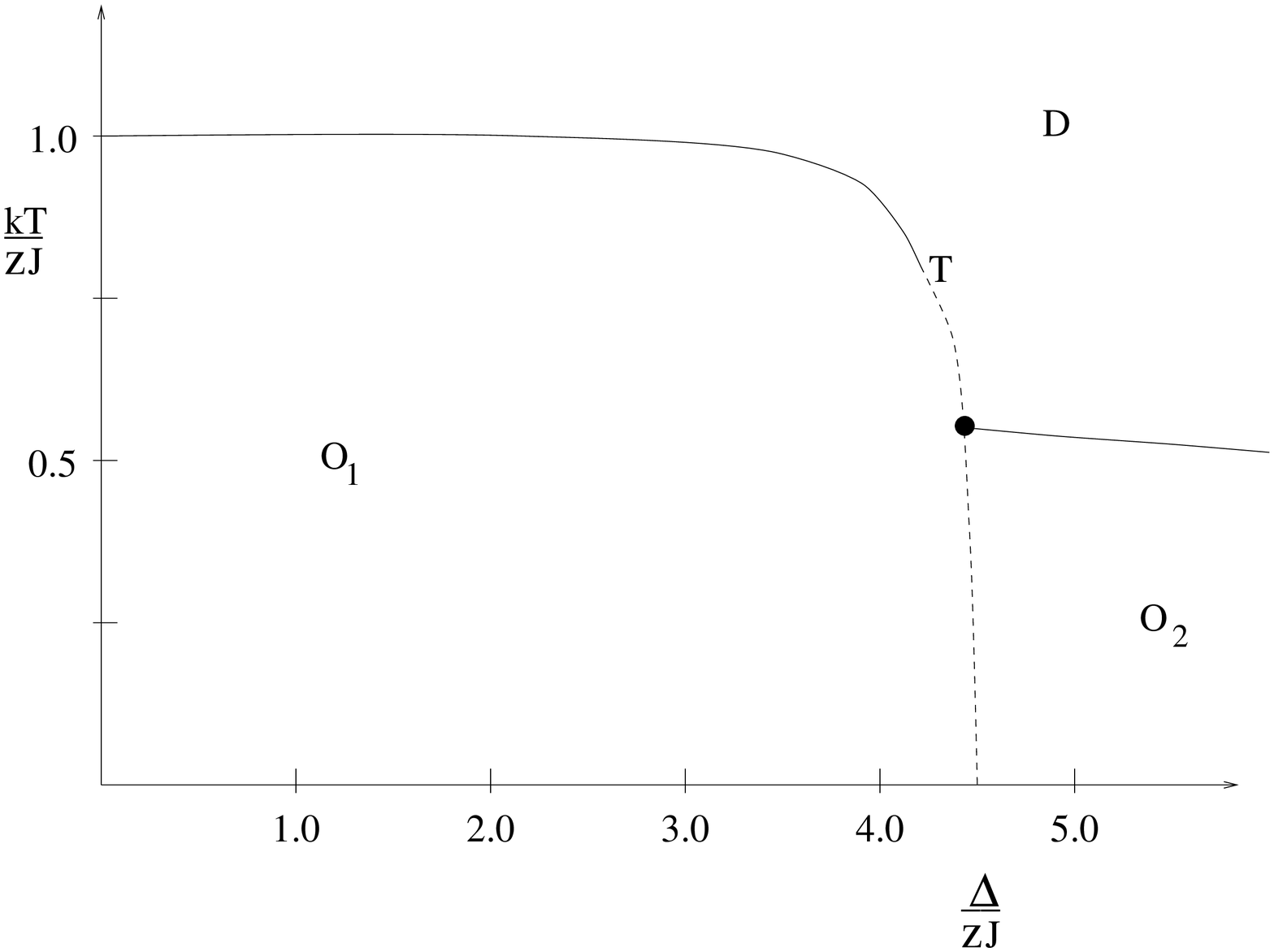}
\caption{Mean-field phase diagram for $K/J=5$ and $r=0.5$.
Same conventions as in Fig.~\ref{k5r03}.}
\label{k5r05} 
\end{center}
\end{figure}

\begin{figure}
\epsfxsize=6.5cm
\begin{center}
\leavevmode
\epsffile{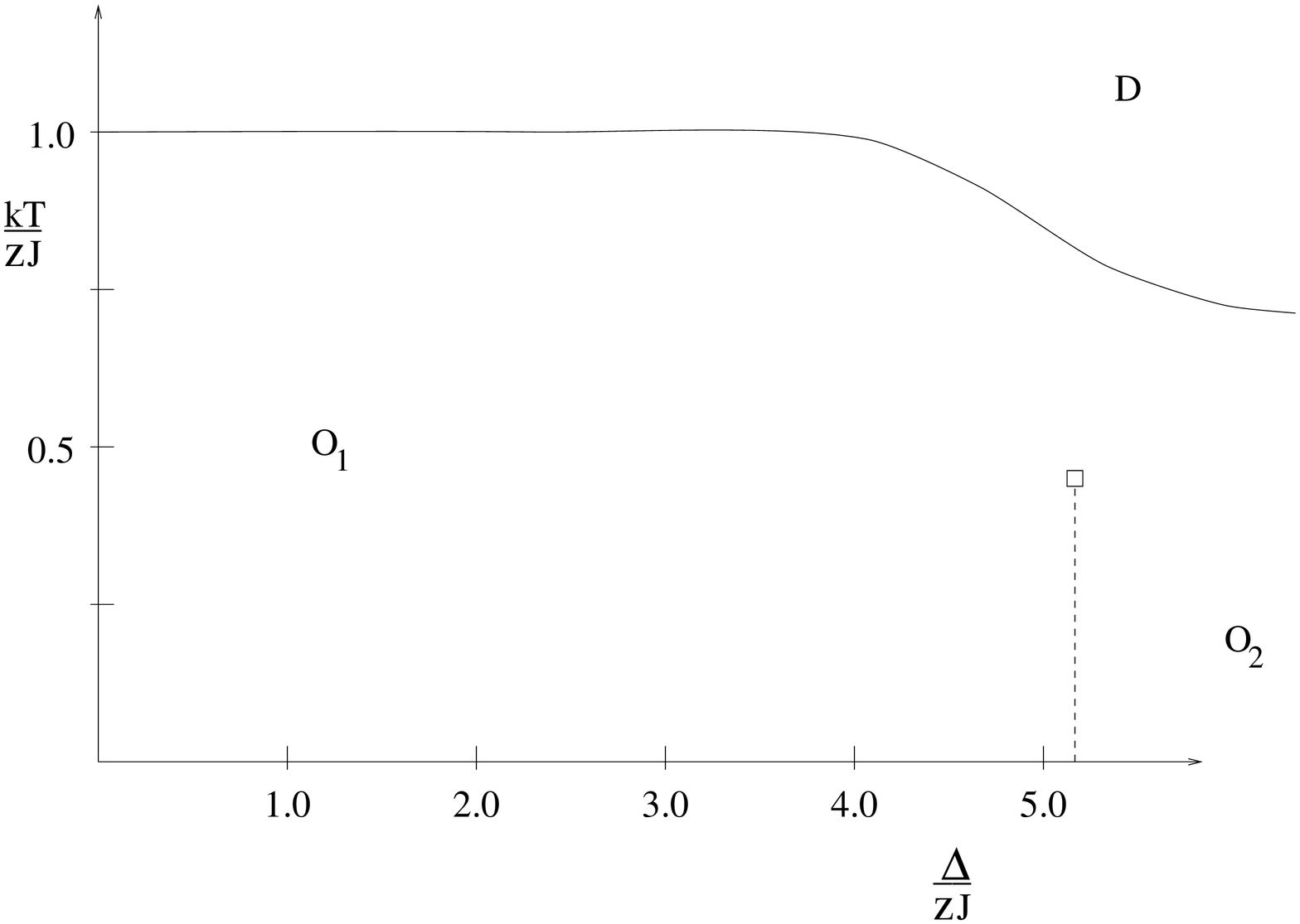}
\caption{Mean-field phase diagram for $K/J=5$ and $r=0.7$.
Same conventions as in Fig.~\ref{k5r03}.}
\label{k5r07} 
\end{center}
\end{figure}

	We now turn to the discussion of the $k_B T / z J \times \Delta / z J $
phase diagrams. In Figs.~\ref{k5r01}, ~\ref{k5r03}, ~\ref{k5r05} and 
\ref{k5r07} we depict sections of 
constant $K/J=5$, for many values of $r$. The phase diagram for $r=0$
(pure BEG model) is qualitatively the same as for $r=0.1$ (Fig.~\ref{k5r01}),
except that the $O_2$ phase is not present. 

	Note that the size of the ordered phases increases with $r$. This is 
expected, since $r$ is the fraction of sites which feel a $-\Delta$ crystal 
field (we have already commented on the ``tail'' which separates the $O_2$ 
and $D_2$ phases, given by $t_c=r$). Another important feature is the presence
of a first-order line between two disordered phases, for $r=0.1$ and $r=0.3$.
In both of these phases $m=0$ but $q>1/2$ for $D_1$, while $q<1/2$ for 
$D_2$. We would like to call attention 
for the phase diagram for $r=0.3$ (Fig.~\ref{k5r03}); this type of diagram is 
not present in the Blume-Capel model.  

	The corresponding phase diagrams for $K/J=3$ show 
only three types of critical behavior: for $r$ near zero,
they are similar to the phase diagram for $K/J=5$ and $r=0.3$
(Fig.~\ref{k5r03}); for intermediate
values of $r$, the behavior is the same as for $K/J=5$ and $r=0.5$
(Fig.~\ref{k5r05}); and
for $r$ close to one, the equivalence is with the diagrams for
$K/J=5$ and $r=0.7$ (Fig.~\ref{k5r07}). 

\begin{figure}
\epsfxsize=6.5cm
\begin{center}
\leavevmode
\epsffile{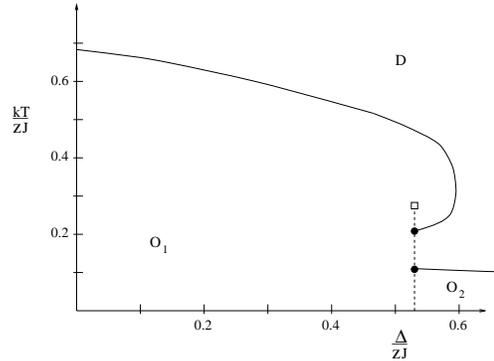}
\caption{Mean-field phase diagram for $K/J=0$ and $r=0.1$. 
Same conventions as in Fig.~\ref{k5r03}.}
\label{k0r01} 
\end{center}
\end{figure}

	The Blume-Capel model ($K/J=0$) has already been studied within
mean-field approximations \cite{mar1,mar2,mf1,mf2}, although for different
probability distributions; the results we find in this case are in qualitative 
agreement with those of Refs.\ \onlinecite{mar2} and \onlinecite{mf1} and
we shall not depict all of them here. The only exception is the diagram
for $r=0.1$ (Fig.~\ref{k0r01}), which is not present for higher values 
of $K/J$.

	On general grounds, one should note that the mean-field approximation
we employed suggests that the random crystal field
does not destroy the first order transitions between disordered
phases and between an ordered and a disordered phase. Even
first order lines between ordered phases (like the one in Fig.~\ref{k0r01})
survive the introduction of randomness.

\section{Two-dimensional Renormalization-group} 
	
	It is well known that mean-field-like approximations are not suitable
to describe low-dimensional systems. We have then to resort to a different
technique, in order to study the RBEG model in two dimensions. RSRG
procedures, on the other hand, have been 
successfully applied to two-dimensional systems. Note, however, that RSRG 
approximations, in general, do
not lead to results as precise as those obtained with Monte Carlo simulations,
phenomenological renormalization or conformal invariance techniques.
Nevertheless, they allow for a correct description of universality classes, 
order of the transitions, crossover phenomena, etc.

	The procedure is the same as the one thouroughly discussed in Ref.
\onlinecite{branco}. There is just one technical point we would like to
stress. Although we start with a uniform distribution for $J$ and $K$,
the renormalization procedure will introduce randomness in all
renormalized quantities ($J', K'$ and $\Delta'$). A possible approach
is to follow the successive renormalized distributions of these
parameters in order to study the phase diagram. We adopted an alternative
way, which forces the renormalized distributions to be the same as the
initial ones, but with renormalized parameters, namely,
$ {\cal P}'_{ap}(J) = \delta(J-J')$, ${\cal P}'_{ap}(K) = \delta(K-K')$ and
${\cal P}'_{ap}(\Delta_i) = r' \; \delta(\Delta_i+\Delta') + (1-r') \;
\delta(\Delta_i-\Delta')$. The values of $J'$ and $K'$ are
obtained imposing that the first moment of the actual distributions for
$J$ and $K$ and of ${\cal P}'_{ap}(J)$ and ${\cal P}'_{ap}(K)$ are equal,
respectively. The values $r'$ and $\Delta'$ are calculated imposing 
that the two lowest moments of ${\cal P}'_{ap}(\Delta)$ match those of the 
real distribution.
This procedure has to be used with some care:
in certain systems, where the random-field mechanism is important and the
initial randomness is on the interaction ($J$, for instance), forcing the field 
back into a uniform distribution leads to incorrect results.
In Ref. \onlinecite{yeo}, for instance, the crystal field probability
distribution is maintained uniform throughout the renormalization
procedure. Consequently, the critical behavior of the random model is 
characteristic of a high-dimensional system: the critical temperature
of the tricritical point diminishes as randomness is increased but only
reaches the zero temperature axis at a finite value of the disorder.
As discussed in Ref. \onlinecite{berker}, the mechanism
responsible for the lack of first-order phase transitions in two-dimensional
random systems is the disorder in the crystal field, which is not taken into 
account by approximations such as the one used in Ref. \onlinecite{yeo}. In
the model we study in this paper, however, the disorder in the field is 
not approximated away by our RSRG
procedure.
	
\begin{figure}
\epsfxsize=6.5cm
\begin{center}
\leavevmode
\epsffile{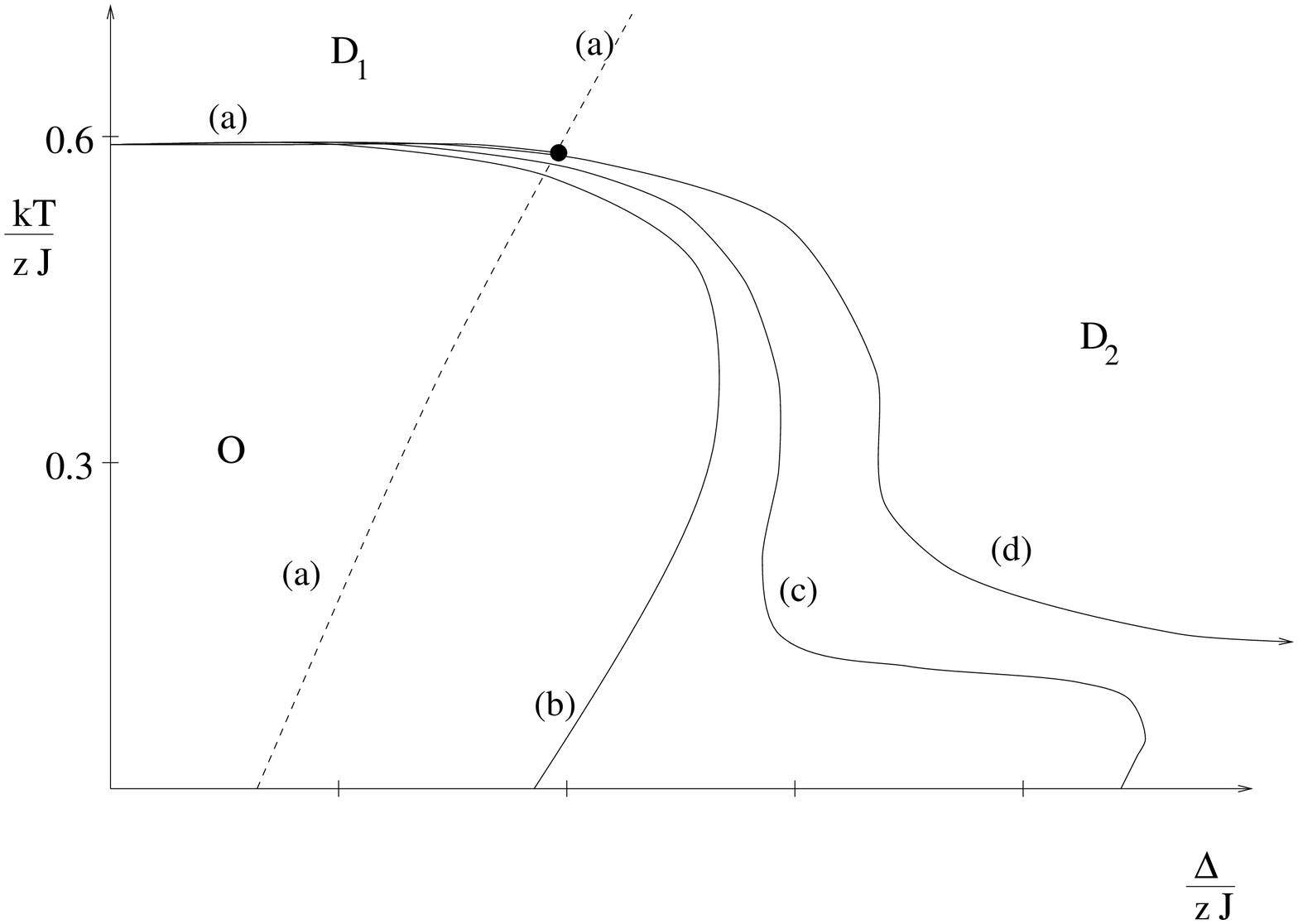}
\caption{Renormalization-group phase diagram for $K/J=2$ and (a) $r=0$, 
(b) $r=0.2$, (c) $r=0.45$, and (d) $r=0.5$. Filled circles stand for critical 
end points, $O$ for
the ordered phase and $D_1$ and $D_2$ for the two disordered phases (see
text). Continuous (dashed) lines represent continuous (first-order)
transitions. The transition lines extend to  $\Delta \rightarrow \infty$ 
only for $r \geq 0.5$. Note that the
critical end point (filled circle) is present only for $r=0$.}
\label{k2r} 
\end{center}
\end{figure}

	Our results for $K/J=2$ are presented in Fig.~\ref{k2r}, where we
depict the $kT/zJ \times \Delta/zJ$ phase diagram for $r=0$ (pure BEG model), 
$r=0.2$, $r=0.45$, and $r=0.5$. 

	Let us first comment on the pure BEG model (curve (a) of
Fig.~\ref{k2r}). As for $K/J=5$ in three dimensions, there are
two types of disordered phases: both have $m=0$  but
$q>1/2$ for phase $D_1$ and $q<1/2$ for phase $D_2$.
The continuous line between
phases $O$ and $D_1$ belongs to the universality class of the
Ising model: this line is attracted to the Ising fixed point,
$C^* \equiv \left( J=0.4407, K=-0.07308, \Delta=-\infty \right)$. 
The dashed line
between phases $O$ and $D_2$ is attracted to the fixed point
$F_1 \equiv \left( J=\infty, K=\infty, \Delta=2(J+K) \right) $, which 
represents
a first-order transition in both $m$ and $q$, i.e., the largest
eigenvalue of the even and the odd sectors of the linearized RGT matrix
are equal to $b^d$ (see Ref. \onlinecite{fisher}). On the other
hand, the dashed line between phases $D_1$ and $D_2$ is attracted to
the fixed point $F_2 \equiv \left( J=0, K=\infty, \Delta=2K+\ln 2 \right)$; 
in this
fixed point only the largest eigenvalue of the {\it even} sector of
the linearized RGT matrix is equal to $b^d$; this is a sign
of a discontinuity in $q$ (but not in $m$) when the line is crossed
\cite{fisher}. 

	In curves $(b)$, $(c)$, and $(d)$ of Fig.~\ref{k2r} we depict the 
$kT/zJ \times \Delta/zJ$ phase diagram for $r \neq 0$. We note that
the first-order line is either replaced by a line of
continuous transitions (between $O$ and $D_2$ phases) or is
eliminated (between $D_1$ and $D_2$ phases), for any infinitesimal
amount of randomness. In fact, the first-order
fixed point attractors, $F_1$ and $F_2$, are unstable along the $r$
direction. There is still a line separating the two disordered phases
(not depicted in Fig.~\ref{k2r}),
$D_1$ and $D_2$, for $r \neq 0$, but this line is attracted to the
$\left( r=1/2, J=0, K=0, \Delta=\infty \right)$ fixed point. This
point represents a model with independent spins, in which no phase
transition can take place.  We note that our results are in accordance with
general arguments on the effects of randomness on multicritical phase
diagrams \cite{berker}, although, to the best of our knowledge, some
features of these arguments have never been tested so far.

	On the other hand, the whole line of continuous
transitions for $r \neq 0$ belongs to the pure Ising model universality class, 
i.e., $C^*$ is a stable fixed-point along the $r$ 
direction. This is expected, since, 
for the hierarchical lattice used in this work, the specific heat
critical exponent of the pure Ising model, $\alpha$, is negative
and disorder is irrelevant, according to the Harris criterion \cite{Har}.
For the corresponding model
on a two-dimensional Bravais lattice, where $\alpha = 0$, the Harris
criterion is inconclusive. The accepted behavior, when disorder is present,
is the following: critical exponents of the random model retain the same
values as their pure conterparts but logarithmic corrections are
introduced by randomness \cite{Fabio}. 
Experimental results also indicate the same critical exponents for
pure and random two-dimensional Ising model \cite{exper}.

	We would like to call attention to the behavior of the critical point 
which separates the $O$ and $D_2$ phases at $T=0$. For $r < 0.5$, 
the transition
at zero temperature takes place at a finite value of $\Delta/zJ$. For
$r \geq 0.5$, the critical line between the ordered and the disordered phases
extends to $\Delta/zJ = \infty$ in the diagram.
In fact, for $\Delta/zJ = \infty$
the RBEG model is equivalent to the site-dilute spin-1/2 Ising model, as
discussed above. Thus, only for
high enough $r$ an infinite cluster of $S=\pm 1$ states will form and will be 
able to sustain order. There is a critical value of $r$, $r_c$, such that
the critical line between the ordered and disordered phases only reaches 
$\Delta/zJ = \infty$ for $r \geq r_c$. 
Our evaluation of $r_c$ is 1/2, while the accepted value for the 
site percolation critical probability on the square lattice 
is $r_c=0.5927$ \cite{perco}. This difference is due to the small cell
we use in this work; nevertheless, the correct qualitative picture is
obtained, i.e., a finite value of $r_c$. 

	Finally, we would like to stress that there are only two types
of phase diagrams for the BEG model; for high values of $K/J$ these
diagrams have the same structure as for $K/J=2$. For small values of
$K/J$, the structure is the same as for the Blume-Capel ($K=0$) model.
As this model has been studied elsewhere \cite{branco}, we will not
discuss it here.

\section{Summary}

       We studied the BEG model in two and three dimensions within
a RSRG framework and a mean-field approximation, respectively. The
disorder is on the crystal field term, which follows a
probability distribution given by: 
${\cal P}(\Delta_i) = r \; \delta(\Delta_i+\Delta) +
(1-r) \; \delta(\Delta_i-\Delta)$. 

	For the mean-field approximation (expected to represent the qualitative
behavior of three-dimensional systems), the presence of randomness 
increases the ordered phase and brings qualitative changes to the 
$kT/zJ \times \Delta/zJ$
phase diagram. More specifically, first-order transitions are present in
the disordered model, but new multicritical points emerge, depending on the
value of $r$.

	In two dimensions, the RSRG approach we use shows that
randomness has a drastic effect on critical behavior:
it supresses non-symmetry-breaking first-order transitions and replaces
symmetry-breaking discontinuous transitions by continuous ones. These
results are in accordance with general arguments concerning the
effects of quenched impurities on multicritical behavior (but, to the best of
our knowledge, the disappearance of the first-order line between disordered 
phases or between ordered phases has never been seen in an actual calculation).
The line
of continuous transition, present for the disordered $(r \neq 0)$ model,
belongs to the universality class of the two-dimensional {\it pure}
Ising model; this results agrees with the Harris criterion, since
the specific heat critical exponent, $\alpha$, is negative for the
hierarchical lattice used in this work. It has been conjectured that a
new unstable critical point, at finite temperatures, might be present
for the disordered system \cite{mar1}; we found no evidence of this point,
for any value of $K$.

	We have also studied the so-called degenerate Blume-Emery-Griffiths
(DBEG) model, introduced in the study of martensitic transitions \cite{barce}.
In the DBEG model, the $S=0$ states are assumed to have a degeneracy $p$, which
mimics the effects of vibrational degrees of freedom. It has been shown
in Ref.\onlinecite{barce} that the effect of increasing $p$ is to shrink
the ordered phase and to increase the region where the transition is of
first-order. Using the same probability distribution for the crystal field as
in the RBEG model, we were able to show that the parameter $p$ may bring
only {\it quantitative} changes to the phase diagrams, for any $K/J$, $r$,
and $p$. This is expected, since the DBEG model is equivalent to the usual
$BEG$ model with all crystal fields displaced by $\ln(p)$ \cite{referee}.
In particular, any infinitesimal amount of randomness in two dimensions
destroys the first order transitions, no matter the value of $p$.

	Finally, we would like to stress that our approximation does not allow
for a study of the BEG model with negative $K$, where new and interesting
critical behavior emerges \cite{ber2}. Work is now being made to discuss this 
model in the presence of a random crystal field.

\section{Acknowledgments}

   We would like to thank Prof. J. F. Stilck for a critical reading of
the manuscript and Prof. Anna Chame for calling our attention to Ref.
\onlinecite{barce} while this work was in progress.

\end{document}